\documentclass[preprint,12pt]{elsarticle}

\usepackage{graphicx}
\usepackage{amsmath}

\usepackage{lineno}

\journal{International Journal of Electrical Power and Energy Systems}

\begin{document}

\begin{frontmatter}


\title{Real Time Impact Control on Charging Energy Storage For Shipboard Power Systems}



\author[label1]{Yusheng Luo}
\author[label2]{Sanjeev Srivastava}
\author[label1]{Manish Mohanpurkar}
\author[label1]{Svetomir Stevic}
\author[label1]{Rob Hovsapian}
\address[label1]{Idaho National Laboratory\\Idaho Falls, Idaho 83415}
\address[label2]{Siemens Corporate Technology\\755 College Rd E, Princeton, NJ 08540}

\begin{abstract}
Medium voltage direct-current based integrated power system is projected as one of the solutions for powering the all-electric ship. It faces significant challenges for accurately energizing  advanced loads, especially the pulsed power load, which can be rail gun, high power radar, and other state of art equipment. Energy storage based on supercapacitors is proposed as a technique for buffering the direct impact of pulsed power load on the power systems. However, the high magnitude of charging current of the energy storage can pose as a disturbance to both distribution and generation systems. This paper presents a fast switching device based real time control system that can achieve a desired balance between maintaining the required power quality and fast charging the energy storage in required time. Test results are shown to verify the performance of the proposed control algorithm.
\end{abstract}

\begin{keyword}
Medium voltage direct-current based integrated power system \sep Pulsed power load \sep Power quality \sep Disturbance metric \sep Real time control


\end{keyword}

\end{frontmatter}


\section{Introduction}
Research related to navy shipboard power system raise a critical concern regarding to the system stability due to diverse loads. Similar to microgrids, navy shipboard power systems do not have a slack bus \cite{ciezki2000selection}. It can be viewed as a microgrid always operating in islanding mode. Compared with typical terrestrial microgrid, the ratio between the overall load and generation is much higher \cite{kondratiev2011invariant}. Although new avenues such as zonal load architecture \cite{doerry2006zonal}, high energy/power density energy storage systems \cite{mo2016hybrid}, and high efficiency generator \cite{doerry2009next} have been developed for supporting stability control, new challenges also arise due to the diverse demand from load side.

In order to reduce ship's size and visibility in future battlefields, an Integrated Power System (IPS) based all-electric ship is developed \cite{mccoy2000powering}. All electric ship is characterized by using electrical motor driven propulsion system to replace traditional mechanical transmission propulsion system. Therefore, the propulsion system is integrated with the shipboard power system, and is termed as IPS. The application of electrical propulsion waives the need of slow speed gear providing power to mechanical propulsion. The remaining fast gear can drive both the generation and electrical propulsion \cite{young2001beyond}. Therefore the space occupied by the transmission system is largely reduced. Nevertheless, the electrical propulsion has high efficiency, which means less fuel is needed and the ship size can be further reduced. Not only the shape of the ship is optimized, but the weapons system can also be improved significantly. Phase-array radars are capable of providing detection with higher resolution and wider area, pulsed power load features in fast, powerful, and accurate attacks, electromagnetic ejector can inject high launch dynamics for shipboard aircrafts and so on. These state of the art devices share a common characteristic, their power demand appears as a large pulse due to the integration of power electronics interface \cite{panda2017design}. The electrical propulsion and Pulsed Power Load (PPL) place a huge burden on the generation and distribution capacity of the shipboard power systems.

To address the challenge of reliably powering the all-electric ship, several efforts have been made. Different distribution architectures such as Medium Voltage Alternative Current (MVAC), High Frequency Alternative Current (HFAC), and Medium Voltage Direct Current (MVDC), have been analyzed \cite{doerry2009next,soman2012model}. Such research efforts reached a conclusion that MVDC system is projected as the optimal distribution architecture for IPS. This is on account of MVDC's no concerns regarding generator synchronization, harmonics distortion, and frequency stability. The electrical devices and related control strategy for supporting MVDC distribution architecture have also been developed \cite{castellan2018review}.

Besides developing IPS, contributions have also been made to test the performance of IPS in powering advanced loads. Propulsion loads occupy a large portion of the total shipboard load consumption. In the rest of the loads, PPLs are usually critical due to their indispensable role in the battle. One of the most challenging issues in controlling IPS operation is providing expected power and energy to PPL without affecting the operation of other critical loads. The impact from PPL to IPS with various distribution architectures are analyzed  in \cite{cassimere2005system,steurer2007investigating,luo2012power}. Analysis shows that the impact from energizing PPL with around 100 MW power can seriously influence the system stability and power quality. In \cite{crider2010reducing}, noteworthy work is presented for an innovative application of power electronics interfaced Supercapacitor Energy Storage System (SESS). Energy storage has found wide application in such hybrid energy systems, for augmenting limited generation and modern loads  \cite{omar2012voltage,chen2016operations}. From the beginning of the development of IPS, energy storage was used as auxiliary power supply  \cite{amy2002considerations}. Supercapacitor is well known for its high power density, which can enable fast charging/discharging \cite{de2013modelling,luopesgm}. By switching the power source from shipboard generation to SESS, the direct impact from PPLs to IPS can be eliminated. However, the charging of SESS can induce high current and still destabilize the IPS. Control methods for driving the SESS charging circuit within desired limits and rates are proposed. In \cite{cassimere2005system}, limit-based supercapacitor control is proposed. Although this method can be effective, it requires a precise understanding of the system's current and power limits in order to prevent abrupt disturbance to the system. In \cite{bash2009medium}, a trapezoidal-based control of SESS charging is introduced. This control is only effective when trapezoidal load profile fits the system requirement, and vice versa. Profile-Based Control (PBC) in \cite{crider2010reducing} features a minimal impact to system stability within a fixed time frame, however it still requires the prerequisite information about the system's current limit. Furthermore, it is challenging to set a fixed charging time which is essential for such systems. Some other papers, such as \cite{im2016cooperative,tan2016adaptive} focus on coordinating SESS charging control and generation control to mitigate the possible disturbance caused by SESS charging. However, during mission time, especially battle time, the quality of communication supporting coordination may not be guaranteed due to weather or battle damage. Nevertheless, these coordination strategies are mostly applied to MVAC system, which highlight coordination of generation control and other load control. The MVDC system is proposed to waiver the concerns of generation control coordination. In another word, it is more open decentralized control strategy. Henceforth a decentralized control strategy is considered as more suitable method for SESS charging in MVDC system.  

The proposed work in this paper is based on the contribution in \cite{luoapplication}, which proposed a Disturbance Metric Control (DMC). DMC is a decentralized technique  developed to control SESS charging according to disturbances monitored in real time from SESS charging in MVDC system. However, it is needed to prove that existing hardware is capable of realizing the proposed real time control theory. To achieve the expected performance of DMC, this study integrates the cutting edge IGBT technology to DMC and proposed Real Time DMC (RTDMC). The paper comprises of the following sections: section 2  describes the experimental system; section 3 analyzes the disturbance due to ES charging; section 4 presents DMC strategy guided by two major concerned disturbance metrics in MVDC system; section 5 describes the hardware implementation of proposed RTDMC strategy; in section 6 test results based on the proposed RTDMC are discussed; and the conclusion is summarized in section 7.
\section{System description}

\subsection{Test Environment}
Unpredicted and unaccounted charging/discharging of SESS can damage hardware leading to failures, making it uneconomical to test in real-world applications. Hence, a digital real time simulator (DRTS) is used to assess the performances and generate result. In this environment, the time-step for simulating electrical circuit is 50 $\mu$s. For power electronics simulation, the time-step can be reduced to 1-2 $\mu$s. The real time computation speed of the power systems can test whether the proposed control strategy is fast enough to adjust SESS charging so that expected control performane can be achieved.

\subsection{MVDC System}

\begin{figure*}[!ht]
\centering
\includegraphics[width=0.9\linewidth,height=0.38\textheight]{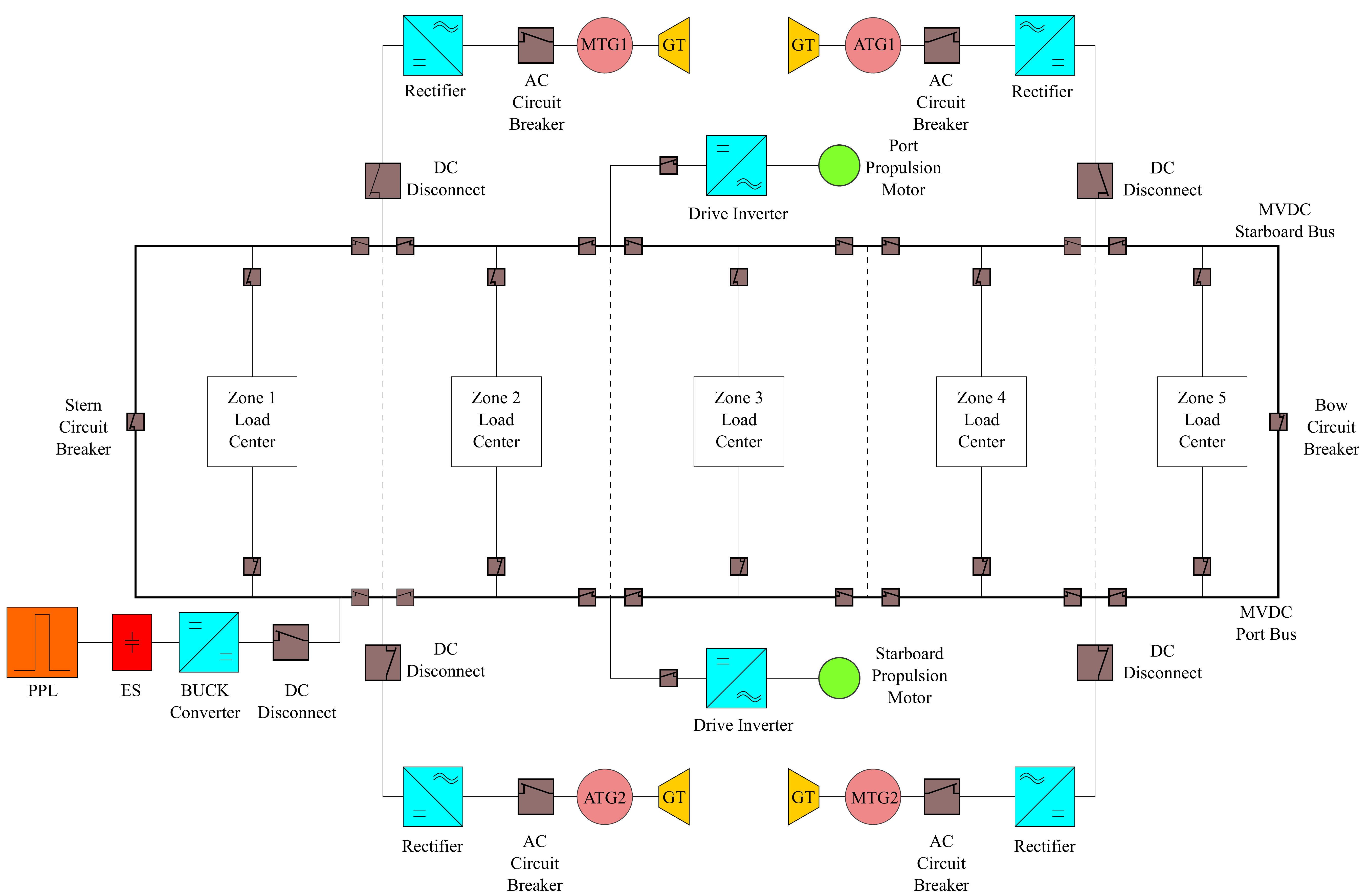}
\caption{System topology of MVDC based IPS.}
\label{MVDC}
\end{figure*}

With the maturity of technology in DC circuit breakers \cite{yeap2018capacitive}, DC cables \cite{sutton2017review}, MVDC system is deemed as advantageous for power distribution. Following the introduction of MVDC in \cite{soman2012model}, we simulated and studied a shipboard IPS with MVDC distribution architecture as shown in Figure ~\ref{MVDC}. In this topology, the main power source are two Main Gas Turbine Generators (MTGs). Each MTG has a generation capacity of 36 MW. In case of need, each MTG is equipped with a 5 MW Auxiliary gas Turbine Generator (ATG). Generation output from MTGs and ATGs is fed into a controllable rectifier and then into a 5 kV MVDC distribution bus. Two 5 kV distribution buses constitute the MVDC distribution system. Since a ship is equally divided into two parts i.e., starboard and port, the bus that feeds starboard is called starboard bus whereas the other one is called port bus. Each bus is directly connected with an MTG, an ATG, and a 36.5 MW propulsion motor. Between two buses, there are zonal loads, stern circuit breaker, and bow circuit breaker. Zonal loads can use circuit breaker to choose drawing power from one bus or both buses. Stern and bow circuit breakers are used to connect or disconnect starboard and port buses. When starboard and port bus are connected with each other, generation from all four generators are merged together, the mode for this connection is called Ring Mode (RM). When starboard and port bus are separated, power supplied to each bus can only come from one MTG and one ATG. We call this operation as Split Plant Mode (SPM). The PPL module is directly connected to port bus, which means the power supply for charging SESS in RM is much more than in SPM.

\subsection{Pulsed Power Load Module}
\begin{figure}
\centering
\includegraphics[width=\linewidth]{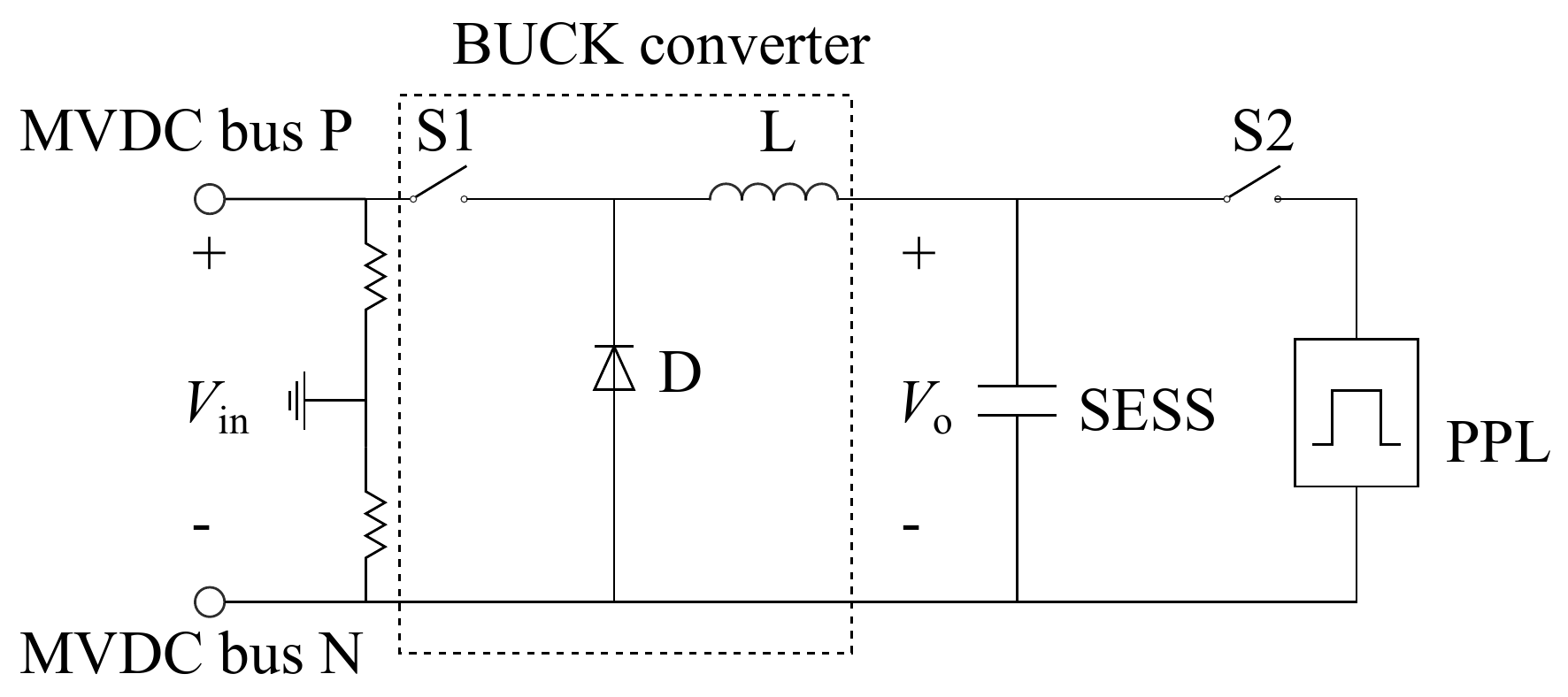}
\caption{Charging circuit for SESS.}
\label{PPL}
\end{figure}

Inside the PPL module, there is PPL, SESS, and the power electronics interface serving as charging circuit of SESS, as shown in Figure ~\ref{PPL}. The charging circuit is a BUCK converter which is controlled by switch S1 \cite{khoudiri2014spectral}. During each commutation period $T$, the turn on time for S1 is $T_{on}$. During $T_{on}$, power goes through S1 and L, then enters SESS. When S1 is opened, current goes through SESS can be continued by passing D and L. When the charging reaches a steady state, the voltage on SESS, $V_o$ can be maintained at a fixed value, as shown in Eq (\ref{buck}):

\begin{equation}
\label{buck}
V_o = \frac{T_{on}}{T}V_{in}.
\end{equation}

During energizing PPL, power is sent from SESS, then S2, finally reaches PPL. To prevent PPL directly draw power from MVDC bus, S1 and S2 are not allowed to be closed simultaneously. Therefore S2 is always opened during charging, and S1 remained opened when PPL is being energized. In this study, totally 300 MJ energy is expected to be injected into SESS. This amount of energy can allows PPL at least be fired up for 2 times without recharging according to \cite{steurer2007investigating}.

\section{Analysis of Disturbance due to SESS charging}

\begin{figure}[!ht]
\centering
\includegraphics[width=0.9\linewidth]{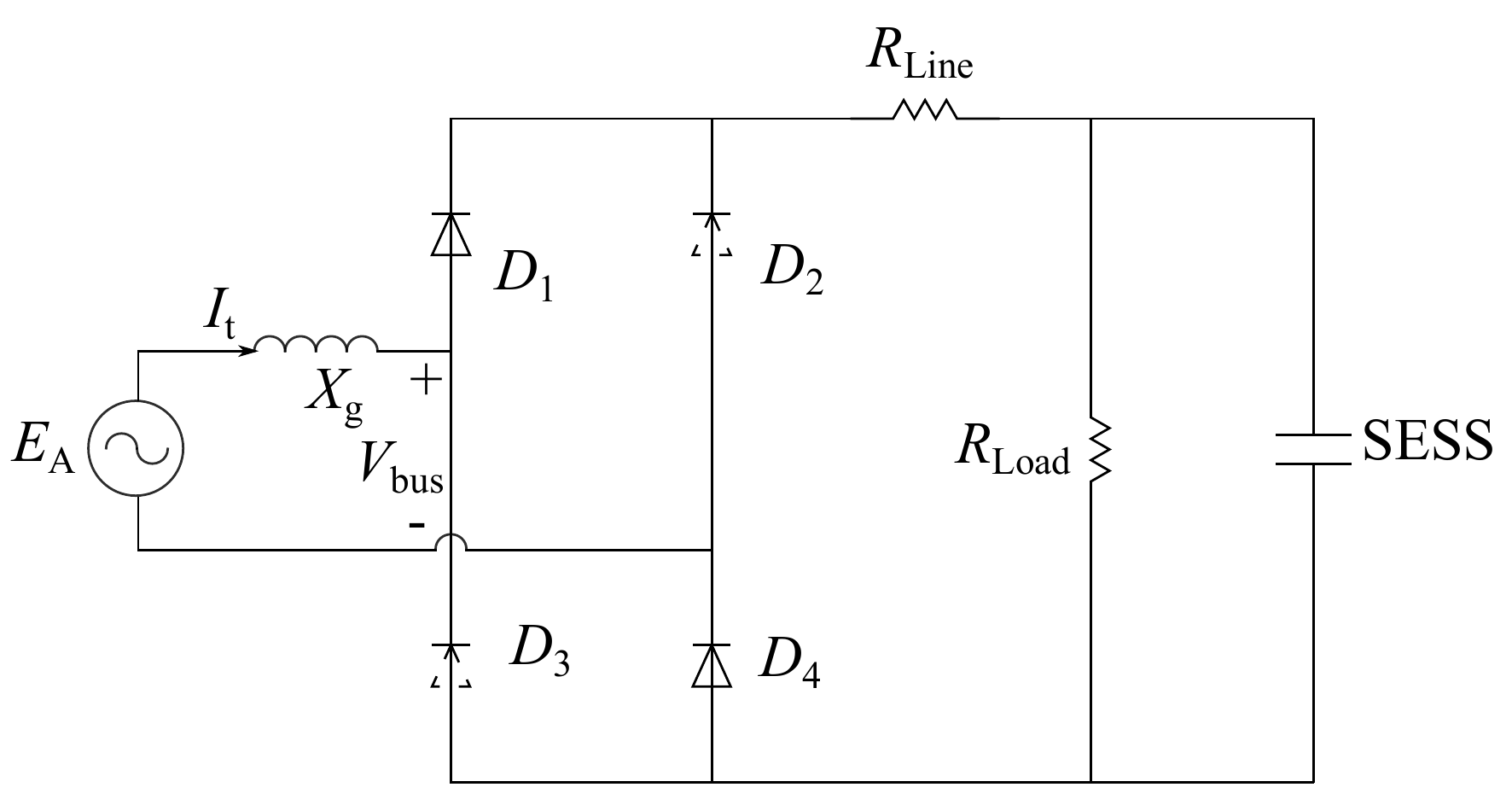}
\caption{Simplification diagram of MVDC based IPS.}
\label{simple}
\end{figure}

Figure ~\ref{simple} is a simplified diagram of the MVDC system using the single line representation, which is used to analyze the possible disturbance caused by SESS charging. In this figure $E_{A}$ is the \emph{emf} generated from a single phase and $X_{g}$ is generator's reactance. The resistance in generator can be ignored during the transient analysis \cite{saadat1999power}. $D_{1}$-$D_{4}$ are the switches forming the rectifier to convert power from AC to DC. $R_{Line}$ is the line resistance in distribution system and $R_{Load}$ is the resistance of loads other than SESS. We set the analysis time begins at $t_0$, when $E_A$ is in the positive semi-period, and only $D_1$ and $D_4$ are turned on. $I_{t}$ is generator's terminal current. Before charging, $I_t$ can be expressed as Eq.(\ref{itori}):

\begin{equation}
\label{itori}
I_{t,t^{-}_{0}}=\frac{E_A}{X_g+R_{Line}+R_{Load}}.
\end{equation}
The terminal voltage which is also the MVDC bus voltage is calculated in Eq.(\ref{vtori})
\begin{equation}
\label{vtori}
V_{bus,t^{-}_{0}}=E_A-X_gI_{Load} .
\end{equation}

Right after that, SESS charging is initiated. Since the voltage of supercapacitor cannot suddenly change, and supercapacitor has zero initial voltage, it is equal to insert a short circuit fault into the system. $R_{load}$ is henceforth removed. $I_t$ calculation is changed to Eq.(\ref{itoai})

\begin{equation}
\label{itoai}
I_{t,t^{+}_{0}}=\frac{E_A}{X_g+R_{Line}}.
\end{equation}
A transient is introduced into the system, the generator stator reactance in Eq.(\ref{itoai}) is replaced by transient reactance $X^{'}_g$ which has smaller value compared with $X_g$, we got Eq.(\ref{itoaii})

\begin{equation}
\label{itoaii}
I_{t,t^{++}_{0}}=\frac{E_A}{X^{'}_g+R_{Line}}
\end{equation}

Compared to (2), both real and imaginary parts of denominator are reduced, the magnitude of denominator is smaller. At this moment, $E_A$'s value can be reviewed as constant, the current magnitude is therefore increased. Since usually $X_g$ is 3-4 times of $X^{'}_g$, and  $R_{Load}>>R_{Line}$, We got the following equation:

\begin{equation*}
\label{vbuschange}
X_g\frac{E_A}{X_g+R_{Line}+R_{Load}}<X^{'}_g\frac{E_A}{X^{'}_g+R_{Line}},
\end{equation*}
which means $V_{bus}$ is reduced.
We can further conclude that the reactive power from generator also increases, according to:
\begin{equation*}
\label{varchange}
X_g(\frac{E_A}{X_g+R_{Line}+R_{Load}})^2<X^{'}_g(\frac{E_A}{X^{'}_g+R_{Line}})^2,
\end{equation*}

Reduced voltage influences the operation of other loads and increased reactive power can reduce the fuel efficiency of generator. When these two disturbances can be mitigated within the tolerated range, SESS charging can be allowed.

\section{Disturbance Mitigation control strategy}

The critical impact is mainly correlated with the charging current, the higher the charging current, the more severe its impact. It is expected that charging control systems must strike a balance between rapid charging and tolerable impact. According to analysis presented in the previous section, power quality impacts from SESS charging are reflected in MVDC bus and generator reactive power output. Two metrics are developed to evaluate the impact from SESS charging:

\begin{enumerate}[leftmargin=*,labelsep=3mm]
\item	$M_1=|V_{bus,lim}-V_{bus}|$, $V_{bus,lim}$ is the input limit of MVDC bus voltage, $V_{bus}$ is the real time measured MVDC bus voltage;
\item	$M_2=Q_{MTG}$, $Q_{MTG}$ is the real time measured generator reactive power.
\end{enumerate}

\begin{figure}[!ht]
\centering
\includegraphics[width=\linewidth]{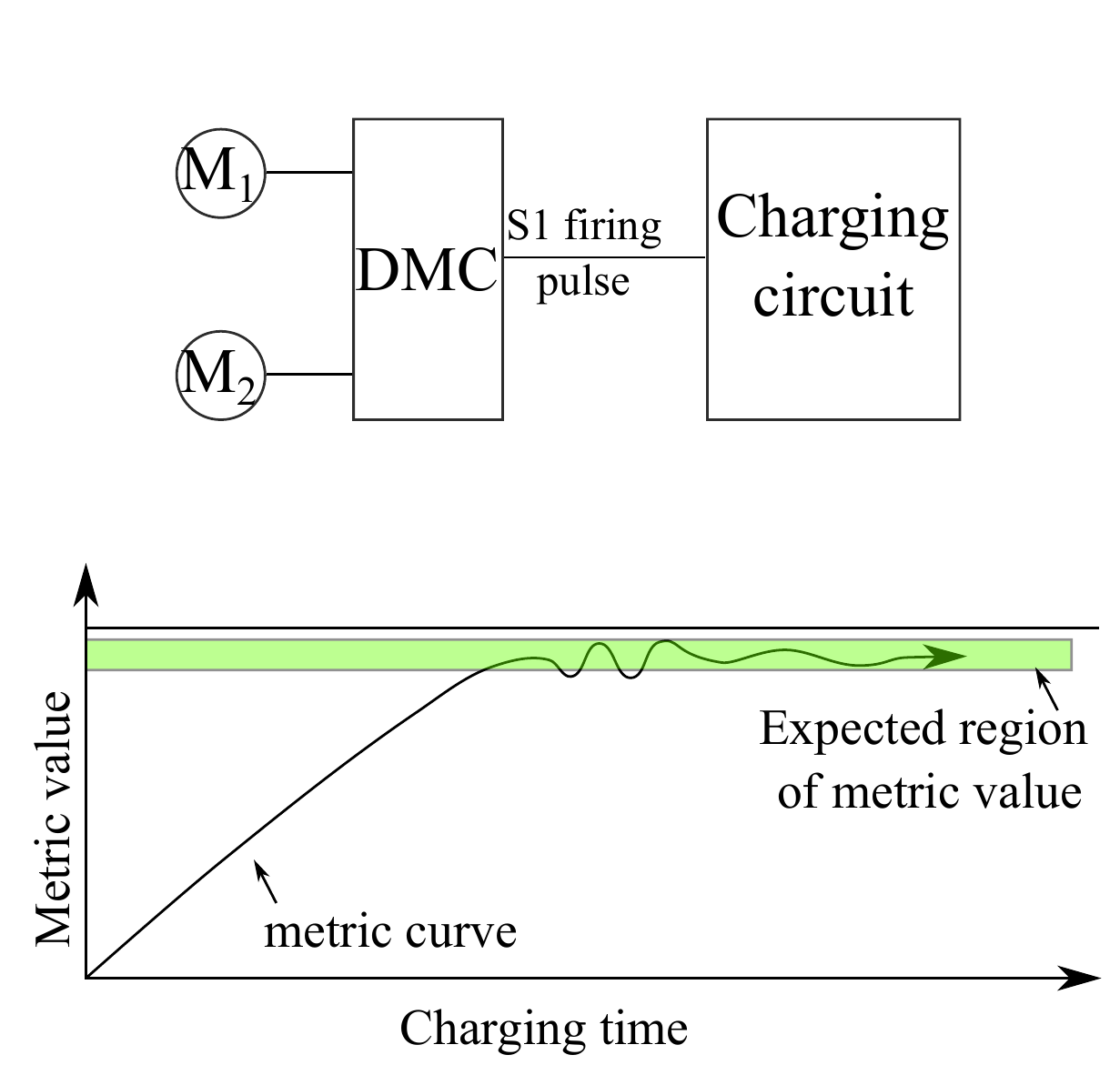}
\caption{DMC control strategy.}
\label{dmc}
\end{figure}

For designing such power systems, standards are employed as guiding principles. One of the applicable standards is related to the expected power quality range and its limitations, such as \cite{5623440}. For any device operating within this system, their control designs should ensure that as long as the specified power quality is maintained, the devices should properly operate as well. Thus, if the SESS charging controller can ensure the power quality is not lowered below the limit required by the standard, all the loads can function normally and their performance is acceptable forming the basis for the DMC proposed in this paper. Assuming the DMC can restrain disturbance below the set limits, the adopters of the proposed work need to modify the limits defined by applicable standards. The customized upper limits of these two metrics are input to DMC, and DMC maintains metric values below limit. Therefore the expected balance can be reached, as shown in Figure ~\ref{dmc}. DMC measures the metric value and determines the control signal sent to charging circuit. The charging current is defined as the current injected from MVDC bus positive node to SESS module. We assume that during the charging process, MVDC bus voltage settle at a steady state other than its rated value. It is expected to maintain charging current with a minor variation, because the charging power can be close to a constant value and facilitate IPS reaching a new steady state during charging. Due to the variations in the characteristics of MVDC bus voltage and MTG reactive power, there are two different procedures for controlling charging current to according to changes of $M_1$ and $M_2$.

\subsection{Control procedure for mitigating impact to MVDC bus voltage}
The charging current can directly affect MVDC power, therefore, if we can maintain charging current at a constant value, the bus voltage can also stay close to a constant value. Hence the key part of mitigating impact to MVDC bus voltage is to find the maximum charging current. The proposed control procedure is:
\begin{enumerate}
\item Set an alert value, which is less than but close to the preset upper limit of bus voltage deviation. 
\item Start the controlled charging process with above alert and metric value.
\item When the metric value reaches the preset alert value, stop charging and record the corresponding charging current.
\item Set this recorded maximum charging current as the reference for charging circuit and let the controller closely tracking it.
\item Stop charging when SESS voltage reaches the desired value.
\end{enumerate}

\begin{figure}[!ht]
\centering
\includegraphics[height=150mm]{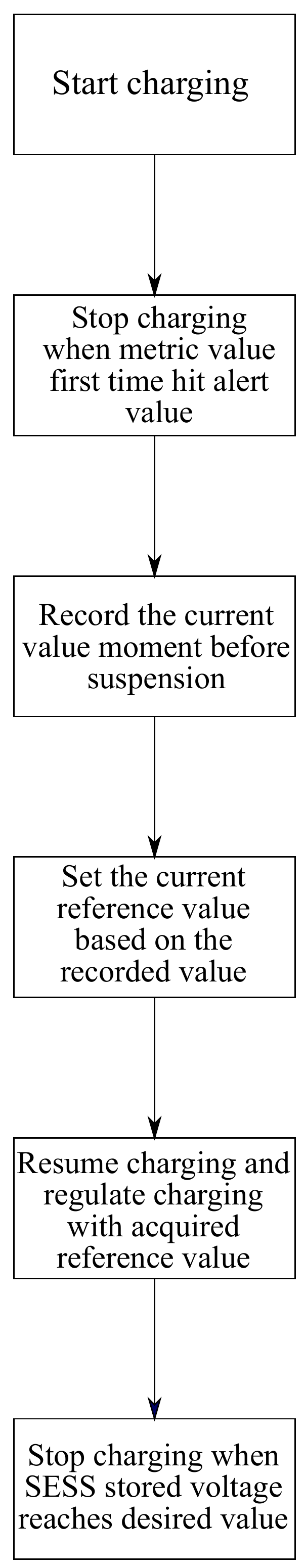}
\caption{Flow chart for mitigating MVDC bus voltage sag}
\label{flowm1}
\end{figure}

\subsection{Control procedure for mitigating impact to MTG reactive power output}
Existing test result shows the impact from charging current to MTG reactive power output is indirect. There is a delay between the charging current change and consequent reactive power change. Therefore, the control procedure in mitigating bus voltage cannot be applied to mitigating reactive power increment. An adaptive control strategy is developed for reactive power output limitation:
\begin{enumerate}

\item Set an alert value, which is less than but close to the preset upper limit of bus voltage deviation. 
\item Start the controlled charging process.
\item When the metric value reaches the preset alert value, stop charging and record the corresponding charging current. Set this charging current value as `maximum charging value' for implementing the algorithm.
\item If the metric value continues to increase and exceeds the upper limit, use a suitable attenuation factor $\alpha$ ($0<alpha<1$) to revise the value of maximum charging value using the formula `$attenuation factorXrecorded charging value$'. 
\item Resume charging and if the charging current increases to the revised maximum charging value, stop charging and monitor the metric change. If the metric value can still reach the upper limit, repeat step 4, 5 and 6.
\item Repeat iteratively until the metric does not exceed upper limit again, set the last value of the product as maximum charging current. Let the upper limit value times a positive coefficient (less than 1) be the lower limit for charging current. This coefficient for the proposed work is set to 0.9.
\item The reference current value is switched between the lower and upper limit based on the actual, monitored value of the charging current. Although the current is oscillating, this oscillation impact can be tolerated due the 10\% difference between upper limit and lower limit.
\item Stop charging when SESS stored voltage reaches desired value.
\end{enumerate}

\begin{figure}
\centering
\includegraphics[height=140mm]{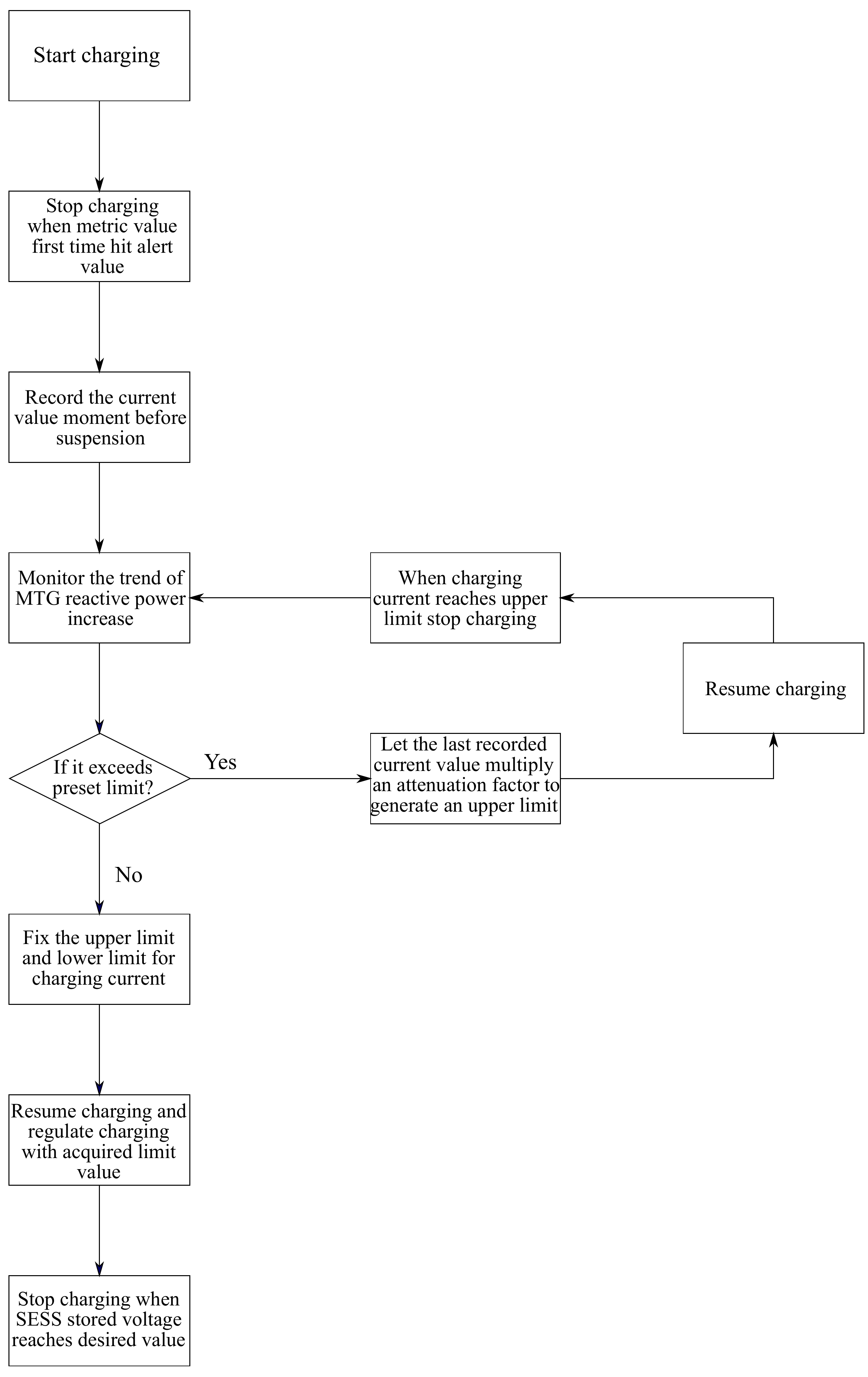}
\caption{Flow chart for limit MTG reactive power output}
\label{flowm2}
\end{figure}

The performance of the proposed control strategy depends on how quickly the critical current value can be captured, how precisely the actual charging current can track the reference current, and how immediately the charging current value can be switched from one state to another state. A desired controller hardware implementation meeting all of the aforementioned prerequisites is expected to be implemented.

\section{Fast Switchable IGBT Based Hardware implementation of Real Time Disturbance Metric Controller}
The expected hardware should be capable of being immediately turning on/off during SESS charging. It should also be capable of withstanding high voltages up to 5 kV during turn off period. Furthermore, the switch should also allow monetary overshoots of charing/discharging current. Based on the these requirements for the fast response of controller, we conducted a literature review related to the fast switchable power electronics devices. It was inferred that a chopper IGBT module FD500R65KE3-K from Infineon has the necessay functionalities for realizing the proposed control strategy. According to \cite{infineon:FD500R65KE3-K}, the turn off delay of this device has been reduced to only 7.3 $\mu$s. It can sustain up to 6.5 kV exerted voltage during turn off. By connecting multiple switches in parallel, a BUCK chopper which is capable of fast turn on/off during high charing/discharing current can be implemented, and the concerned disturbance can be maintained within desired regions. This IGBT module is commercially-off-the-shelf and there may be similar products with the necessary functionalities. The expected energy to be stored in SESS is 300 MW. To reduce the size of the capacitor, the stored voltage should be as high as possible. On the other hand, the highest voltage stored in SESS should not be too high to affect the turning on/off performance of S1 and S2 in Figure ~\ref{PPL}. According to \cite{infineon:FD500R65KE3-K}, the best performance is verified when $V_{CE}$ of the switch device is 3600 V. We allow a 10\% increase of test voltage, and the final charged voltage is set at 4 kV. According to Eq. (\ref{energy}), the capacitance of SESS in this study case is 37.5 F.
\begin{equation}
\label{energy}
E_{c} = \frac{1}{2}CV_c^2.
\end{equation}

\section{Test Result}
In order to validate whether the proposed test can properly mitigate the impact to IPS and ensure fast charging, test case is setup in the DRTS with a simulation time step of 50 $\mu$s. The reaction of test system can well approach the actual real time power system response. Overall initial generation output in all the study cases is set to 70 MW and 5 MVAr. The generation margin left for SESS charging is 30 MVA. 4 test cases are run: $M_{1,limit}$=0.6 kV, $M_{1,limit}$=0.8 kV, $M_{2,limit}$= 6 MVAr, $M_{2,limit}$=10 MVAr. SESS charging begins at the fifth second after a test is initialized.
\subsection{Test case for MVDC bus voltage sag mitigation}
\begin{figure}[!h]
\centering
\includegraphics[width=\linewidth]{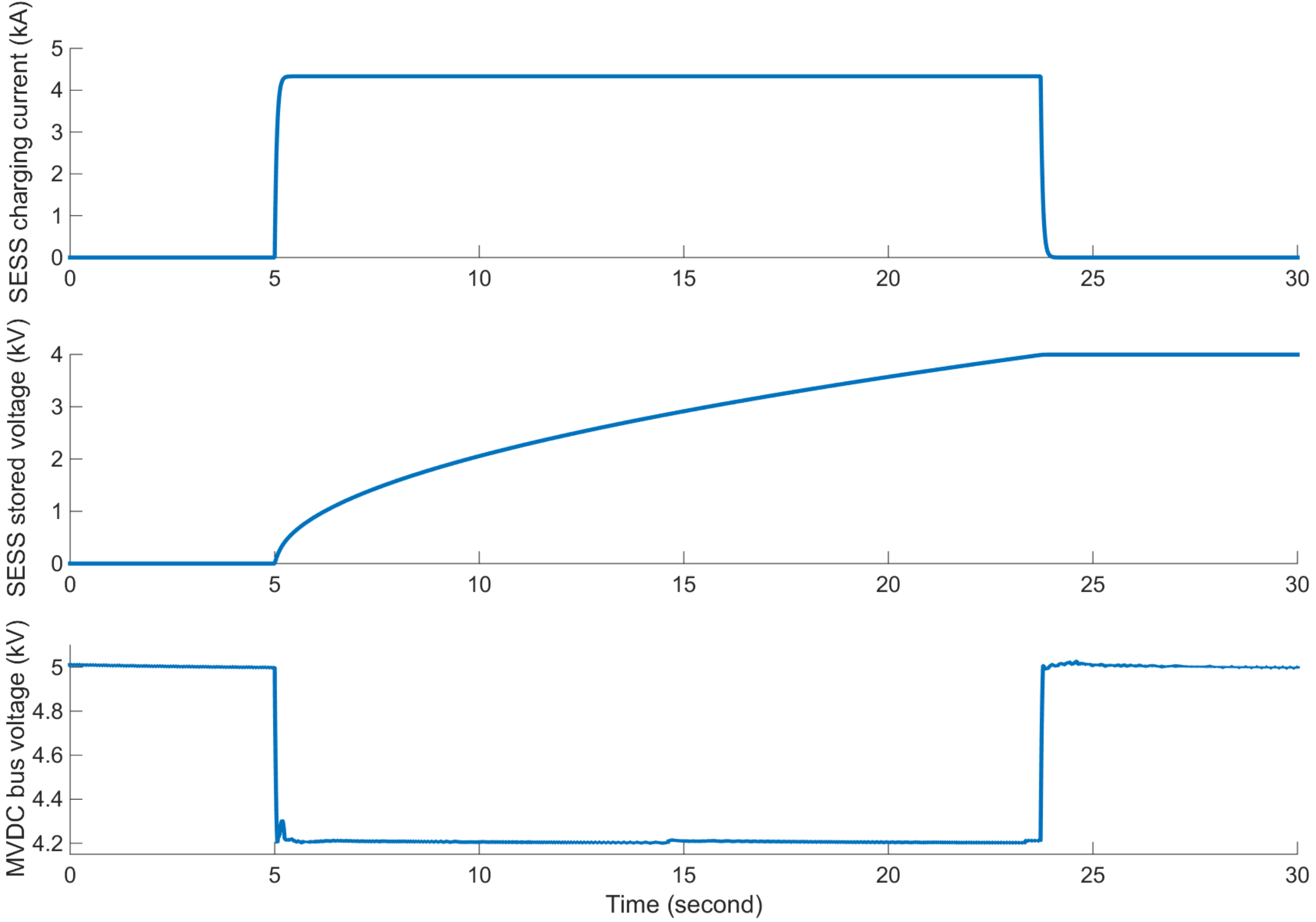}
\caption{Case study: $M_{1,limit}=0.8$ kV}
\label{m108}
\end{figure}

Figure ~\ref{m108} shows test result when $M_{1,limit}$ is set to 0.8 kV and the alert value is set to 4.205 kV. When bus voltage goes down and reaches this value, the charging is suspended and current value is marked as 4.3 kA. Then, charging controller enables the actual current value stay at 4.3 kA. The charging process lasts 19 second and stops when SESS stored voltage reaches 4 kV. The minimum bus voltage is 4.201 kV. Average bus voltage value during charging is 4.208 kV, and the maximum value is 4.3 kV. During the charging the expected bus voltage value is maintained.

\subsection{Test case for MTG reactive power output restrain}
\begin{figure}[!h]
\centering
\includegraphics[width=\linewidth]{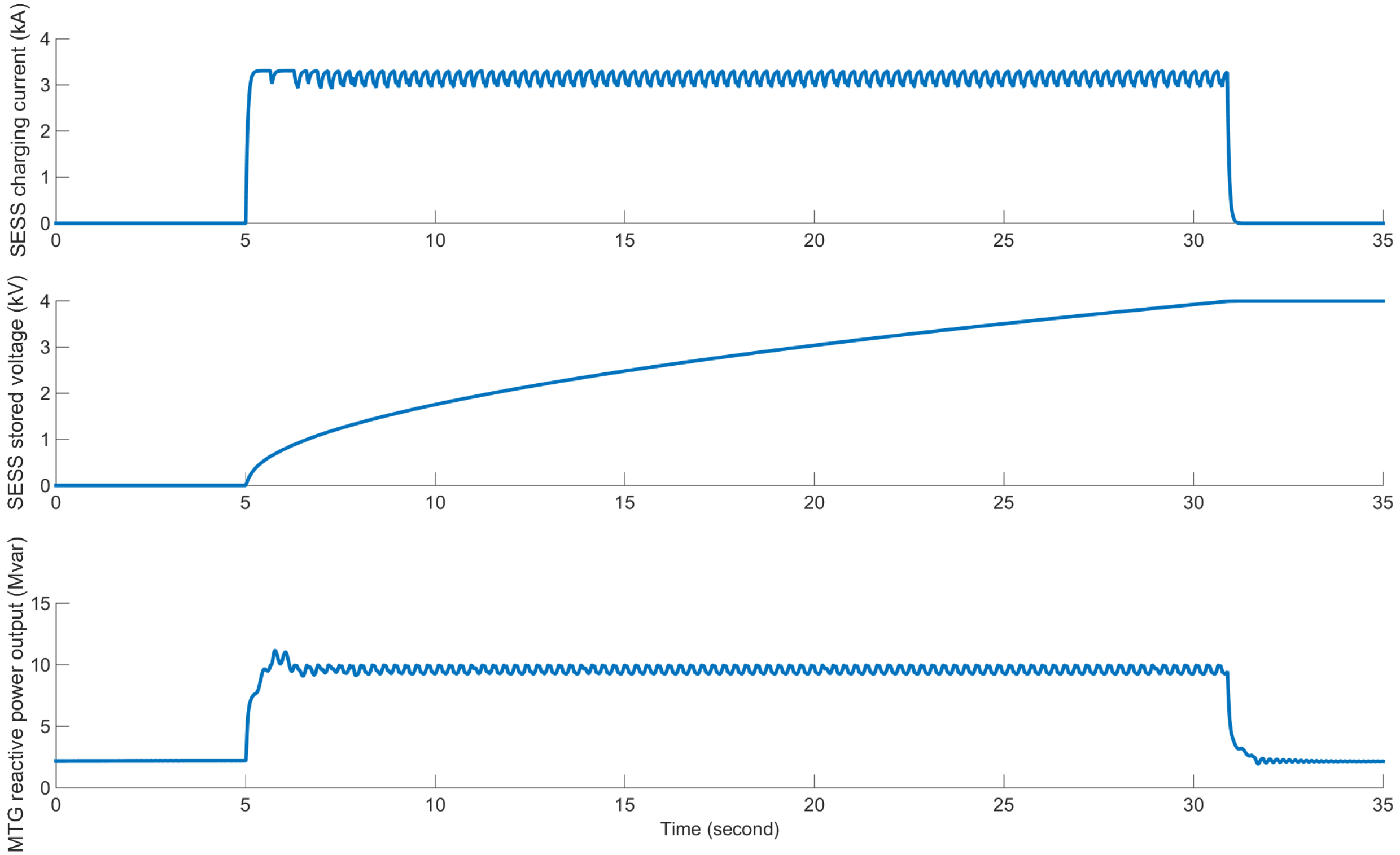}
\caption{Case study: $M_{2,limit}=10$ Mvar}
\label{m210}
\end{figure}

Figure ~\ref{m210} is used to demonstrate the effect of DMC when the limit of $M_2$ is set to 10 MVAr with the alert value is set to 9.5 MVAr. Attenuation factor is set to 0.95. When charging current first reaches 3.3 kA, single MTG reactive power output reaches 9.5 MVAr and hence charging is suspended. MTG output continues to increase and peaks at 9.641 MVAr. Therefore 3.3 kA is set as the upper limit for charging current, and 2.97 kA is set as the lower limit. It can be seen that after the setting, reactive exceeds the limit twice, this is caused by the adjustment of generator excitation. After a short transient, reactive power is stays below 10 Mvar. The average value of $M_2$ is 9.58 MVAr. It takes 26 seconds to inject 300 MJ energy to SESS. Attenuation factor is set to 0.9.

Table ~\ref{table} shows the summarization of all 4 test cases. Test results show the proposed DMC can strike an optimal balance point between fast charging and required power quality acquisition.

\begin{center}
\begin{table*}[ht]
{
\caption{Test cases result summarization}
\label{table}
\resizebox{\columnwidth}{!}{%
\begin{tabular}{llllll}
test setting & maximum metric value & minimum metric value & average metric value & charging current value & charging time \\
$M_{1,limit}$=0.6 kV & 4.04 kV & 4.401 kV & 4.02 kV & 3.9 kA & 21 s \\
$M_{1,limit}$=0.8 kV & 4.3 kV & 4.201 kV & 4.208 kV & 4.3 kA & 19 s \\
$M_{2,limit}$=6 Mvar & 6.04 Mvar & 5.84 Mvar & 5.89 Mvar & (1.8, 2.0) kA & 50 s \\
$M_{2,limit}$=10 Mvar & 11.15 Mvar & 9.23 Mvar & 9.58 Mvar & (2.97, 3.3) kA & 26 s
\end{tabular}}%
}
\end{table*}
\end{center}






\section{Conclusions}

In this paper, DMC strategy is proposed based on a real-time analysis of SESS charing in MVDC based IPS. The proposed control strategy focus on reaching a reasonable balance between fast charging SESS and maintaining required power quality. Based on the first stage analysis, the impact from SESS charging to power quality is assessed and studied. The underlying reason of the impact from SESS charging to power quality is the large charging current. It is challenging to quantify the relation between charging current and power quality. One feasible way for limiting impact from charging current is an online real time monitoring charging current and power quality. Based on the real time monitoring result, an optimal charging current value can be generated. The next challenging part is how to drive the charging current rapidly and precisely track the reference value. A fast switching charging circuit based on the latest IGBT technology development is proposed. Test results shows that using fast IGBT to implement SESS charging system can accurately follow the control command sent from DMC. Realization of fast SESS charging can maintain the required power quality.

In the future, greater disturbances caused by SESS charging, such as the AC side harmonic influence on MVDC bus, is expected to be analyzed. The proposed control strategy highly depends on the rapid monitoring and action. In this paper, using DRTS, the effect of real time control has been simulated. Simulation result shows if the device for monitoring and controlling is fast enough, rapid SESS charging can be realized without sacrificing required power quality. In the future, hardware-in-the-loop test in real time will be performed in order to validate the proposed control strategy on actual hardware. Finally, decentralized coordination between SESS charging control and generator control will also be studied as future work.

\section{Acknowledgements}
Proposed work is supported by the Water Power Technology Office, Department of Energy and the INL Laboratory Directed Research \& Development (LDRD) Program under DOE Idaho Operations Office Contract DE-AC07-05ID14517.




\bibliographystyle{model1-num-names}
\bibliography{IEEEexample}







\end{document}